\documentstyle[aps,prl,multicol,epsfig,amssymb,amstex]{revtex}

\begin{document}

\title{Magnetically induced chessboard pattern in the conductance of a Kondo quantum dot}

\author{M. Stopa$^1$ and W.G. van der Wiel$^2$}
\address{$^1$ERATO-JST, 4S-308S,
NTT Atsugi Research and Development Laboratories, \\
3-1 Morinosato-Wakamiya, Atsugi-shi Kanagawa-ken, 243-0198 Japan}
\address{$^2$PRESTO-JST; University of Tokyo, 7-3-1, Hongo, Bunkyo-ku, Tokyo 113-0033,
Japan; Department of NanoScience and DIMES, Delft University of
Technology, PO Box 5046, 2600 GA Delft, The Netherlands}

\author{S. De Franceschi$^3$, S. Tarucha$^4$ and L.P. Kouwenhoven$^3$}
\address{$^3$ERATO-JST, Department of NanoScience and DIMES, Delft University of
Technology,\\
PO Box 5046, 2600 GA Delft, The Netherlands}
\address{$^4$ERATO-JST, University of Tokyo, 7-3-1, Hongo, Bunkyo-ku, Tokyo 113-0033,
Japan}

\maketitle

\begin{abstract}
We quantitatively describe the main features of the magnetically
induced conductance modulation of a Kondo quantum dot -- or
chessboard pattern -- in terms of a constant-interaction double
quantum dot model. We show that the analogy with a double dot
holds down to remarkably low magnetic fields. The analysis is
extended by full 3D spin density functional calculations.
Introducing an effective Kondo coupling parameter, the chessboard
pattern is self-consistently computed as a function of magnetic
field and electron number, which enables us to quantitatively
explain our experimental data. \vspace{-0.75cm}
\end{abstract}

\pacs{73.23.-b,73.23.Hk,72.15.Qm}
\begin{multicols}{2}

A quantum dot \cite{Kouwenhoven97} with a finite net electron spin
strongly coupled to its leads, enabling higher-order co-tunneling
processes, is able to exhibit the Kondo effect
\cite{Glazman88,Ng88}. The Kondo effect in quantum dots manifests
itself as an enhanced conductance in the Coulomb blockade regime
and occurs for temperatures and source-drain voltages below an
energy scale set by the Kondo temperature. The first experimental
results on the Kondo effect in quantum dots
\cite{GG98,Sara98,Schmid98} were described with help of the
spin-1/2 Anderson impurity model. Assuming continuous filling of
spin-degenerate single-particle levels, the dot is expected to
either have total electron spin $S=0$ (for electron number $N$ is
even) or $S=1/2$ (for $N$ is odd). The Kondo effect is therefore
only expected for odd $N$, hence giving rise to an
``even-odd-effect" in the Coulomb valley conductance. A wide range
of experiments, however, has shown a clear deviation from this
picture
\cite{Maurer99,Wiel00,Sasaki00,Nygard00,Schmid00,Keller01,Wiel02,Sprinzak02,Fuhner02,Keyser02}.
Particularly striking is the observation of a ``chessboard
pattern" in the dot conductance as a function of magnetic field,
$B$, and gate voltage, $V_g$
\cite{Schmid00,Keller01,Sprinzak02,Fuhner02,Keyser02}.
Characteristic for this pattern is the alternation of high and low
valley conductance regions as a function of $B$ within the same
Coulomb valley, i$.$e$.$ for \emph{constant} $N$. In addition, the
conductance also alternates when $N$ is changed by sweeping $V_g$
at constant $B$. The distinct regions in the $V_g$,$B$ plane of
either high or low conductance are associated with the fields of a
chessboard due to the similar appearance when the conductance is
plotted in color scale (see Fig$.$ 1b). It has been experimentally
shown that the enhanced conductance in certain Coulomb blockade
regions can be ascribed to the Kondo effect, for both $N$ odd
\emph{and} even
\cite{Schmid00,Keller01,Sprinzak02,Fuhner02,Keyser02}.\\
\indent In this paper, we start by presenting our experimental
data on a single lateral quantum dot clearly exhibiting the
chessboard pattern. Next, we calculate the skeleton of the
chessboard pattern with a constant interaction (CI) model of two
capacitively interacting dots, formed by the two lowest Landau
levels (LLs) -- an analogy (partly qualitatively) applied before
at high magnetic fields \cite{Keller01,McEuen92,Kinaret93,Vaart94}
-- and obtain the characteristic hexago-
 \begin{figure}[htbp]
 \vspace{-0.5cm}
   \begin{center}
   \centerline{\epsfig{file=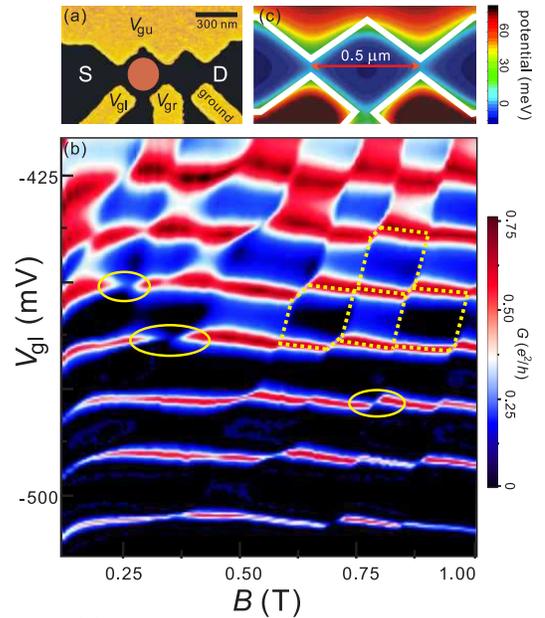, width=7cm, clip=true}}
     \caption{(a) Scanning electron micrograph of the device. The
     metal gates are yellow and the dot is indicated by a red circle. The ungated 2DEG
     has a mobility of 2.3 $\times$ $10^6$ cm$^2$/(Vs) and
     an electron density of 1.9 $\times$ 10$^{15}$ m$^{-2}$ at 4.2 K.
     The nominal dot size is 320 $\times$ 320 nm$^2$. (b) Color scale
     plot of the experimental linear conductance $G$ through the dot as a function
     of $B$ and $V_{gl}$ at 10 mK. The dotted hexagons highlight the shape
     of a few chessboard fields. The ellipses indicate some regions where suppression
     of the Coulomb peak occurs. (c) Calculated self-consistent potential
     landscape of the device. The white lines denote the contours of the metal gates.}
   \end{center}
   \label{fig1}
 \end{figure}
\noindent nal, double dot (DD) stability diagram \cite{Wiel03}.
Interestingly, in this work we find that, due to small $N$ and the
shallow potential, higher LLs only start to fill below a {\it few
tenths of a tesla}. Thus the DD analogy applies down to remarkably
low magnetic fields. We finally perform full, 3D spin density
functional (SDF) calculations for our device and introduce an
effective Kondo coupling, derived entirely from the
self-consistent results. Hence we can simultaneously calculate the
electronic states of the dot and an estimate of the Kondo
coupling, exhibiting the chessboard structure, and quantitatively
explain some of its
subtler features.\\
\indent Our quantum dot is shown in Fig$.$ 1a. Metal gates are
deposited on top of a GaAs/AlGaAs heterostructure with a
two-dimensional electron gas (2DEG) 100 nm below the surface
\cite{Wiel03}. By depleting the 2DEG below the gates, the quantum
dot is defined. Current can flow from the source (S) to the drain
(D) contact. The electron number is varied by sweeping the left
gate voltage, $V_{gl}$. The SDF calculations show that our dot
typically contains $\sim$$20-40$ electrons.\\
\indent Figure 1b shows a color scale plot of the linear
conductance $G$ through the dot as a function of $B$ and $V_{gl}$.
Red (blue) corresponds to large (small) $G$ (see scale in Fig$.$
1b). For the most negative values of $V_{gl}$ the coupling of the
dot to the leads is weak. This results in relatively sharp Coulomb
peaks (red lines) and low valley conductance (dark blue regions).
However, if $V_{gl}$ is increased, the valley conductance reaches
considerable values ($\sim$$e^2/h$) in certain regions of the
($B,V_{gl}$) plane. Most strikingly, the regions of low and high
valley conductance alternate both along the $V_{gl}$ and the $B$
axis in a regular fashion, resulting in the aforementioned
chessboard pattern. The $V_{gl}$ period ($\sim$10 mV) is set by
the energy required for adding an extra electron to the dot
(addition energy), whereas the $B$ period ($\sim$0.1 T)
corresponds to adding a flux quantum to the effective dot area.
Based on the temperature dependence of $G$ in the high valley
conductance regions (not shown here), we can ascribe the
enhancement of $G$ to the Kondo effect. The transition from low to
high valley conductance is associated with an abrupt jump of the
$V_{gl}$ position of the Coulomb peaks. In some cases (see
ellipses in Fig$.$ 1b) the jump is accompanied by a suppression of
the peak height. Below we present a quantitative description for
the experimental features discussed above, using first an
intuitive, though quantitative, approach
followed by a fully self-consistent simulation of our device.\\
\indent In the region where the two lowest LLs, labeled LL0 and
LL1, are occupied, the outline of the chessboard pattern can be
considered to be a DD (Fig$.$ 2a) stability diagram, where now $B$
acts as one of the gates. $B$ couples to both ``dots'' (i$.$e$.$
LLs) but with different ``lever arms.'' Similarly, $V_{gl}$
couples with different capacitances to the two LLs. We write an
elementary CI functional for the total energy, $E$,
\begin{equation}
\begin{split}
&E(N_0,N_1) = \sum_{nm} \sum_{\sigma} \varepsilon_{nm} +\\
&\frac{1}{2D}[C_{11}(N_0e - C_{g0}V_{gl})^2 + C_{00}(N_1e -
C_{g1}V_{gl})^2] \\
&-  \frac{C_{01}}{D}(N_0e - C_{g0}V_{gl}) (N_1e - C_{g1}V_{gl}),
\end{split}
\label{eq:double}
\end{equation}
where $N_{0(1)}$ is the number of electrons in LL0(1),
$\varepsilon_{nm} \equiv (2n + |m| + 1) \hbar \tilde{\omega} + m
\hbar \omega_c$ are the Fock-Darwin (FD) energy levels \cite{FD},
and the sum is over the lowest two LLs. A LL consists of all $n,m$
that satisfy LL index $\lambda=n+\frac{1}{2}(|m|+m)= constant$
(i$.$e$.$ 0 or 1 for our specific case), and $\sigma$ is the spin
index (since at the magnetic fields considered here the Zeeman
splitting is much smaller than the other relevant energy scales,
we assume spin-degenerate states). Here, $\tilde{\omega}^2 \equiv
\omega_0^2 + \omega_c^2$ with $\omega_0$ the bare confining
frequency and $\omega_c$ the cyclotron frequency. $C_{ij}$ are the
capacitance matrix elements (off-diagonal elements are always
negative \cite{Wiel03,Stopa95}), $D \equiv C_{00} C_{11} -
C_{01}^2$. Here $V_{gl}$ is defined relative to the gate voltage
that induces $N_{0(1)}^0$ electrons (i$.$e$.$ $N_{0(1)}^0$ is the
number of electrons in LL0(1) at $V_{gl}$=0).\\
 \begin{figure}[htbp]
\vspace{-1cm}
   \begin{center}
     \centerline{\epsfig{file=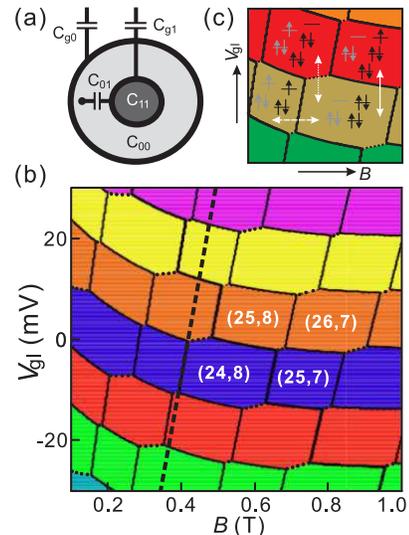, width=5.4cm, clip=true}}
     \caption{(a) Schematic of the dot in terms of two Landau levels (LLs).
     (b) Results of numerical
     minimization of Eq$.$ 1. Color stripes are
     regions of constant $N = N_0 + N_1$. Numbers in
     parentheses show $N_0,N_1$. Black lines bound regions of
     constant $N_0$. Dotted lines show ``$N_1$ boundaries''. The $N_1$ boundary
     lengths alternate (compare to Fig$.$ 1b). Capacitances and dot parabolicity
     estimated for $N_0^0=25$, $N_1^0=10$, $B_0=0.6 \, T$:
     $C_{00}=3.3$, $C_{11}=2.2$, $C_{01}=-2.1$, $C_{g0}=-0.10$ and
     $C_{g1}=-0.056$ (all in aF), $\omega_0 = 0.4 \,
     meV$. The dashed line is the approximate 3rd LL filling boundary.
     (c) Schematic honeycomb structure in the $B$,$V_{gl}$ plane resulting from
     the double dot model. Typical occupancy configurations
     (energy vs. position) of LL1 (gray arrows) and LL0 (black arrows) at the
     dot boundary are shown for different honeycomb cells.}
     \label{fig2}
   \end{center}
    \end{figure}
\indent First, we estimate capacitances $C_{ij}$ and bare
confining potential $\omega_0$ from the full SDF calculations for
$(N_0^0,N_1^0)$ electrons and for a ``central'' magnetic field
$B_0$ (see caption, Fig$.$ 2) \cite{caps}. Then, we numerically
minimize Eq$.$ \ref{eq:double} with respect to $N_0$ and $N_1$,
keeping $C_{ij}$ and $\omega_0$ constant (even though in principle
these change slightly over our ($B$,$V_{gl}$) range). We find a
``honeycomb" structure characteristic for DDs \cite{Wiel03}. As
shown in Figs$.$ 2b and c, this structure clearly emerges and
compares reasonably well with the experimental data of Fig$.$
1b.\\
\indent Figure 2c schematically shows the population of the lowest
two LLs in four neighboring honeycomb cells. $B$ principally
induces depopulation of LL1 to LL0; $V_{gl}$ changes the total
electron number $N=N_0+N_1$. Note that the dotted boundaries
between $N_1$ and $N_1 +1$ (``$N_1$ borders''), are considerably
shorter than the solid boundaries between $N_0$ and $N_0 + 1$.
Weak {\it tunnel} coupling of the inner LL (LL1) to the leads
causes these $N_1$ borders to appear as gaps (or small offsets) in
the Coulomb oscillations \cite{Ciorga02}, which are clearly seen
in Fig$.$ 1b. Analysis of Eq$.$ \ref{eq:double} shows that the
length of the $N_1$ borders, in both $B$ and $V_{gl}$, is
proportional to $C_{11}-|C_{01}|$. Therefore, the short $N_1$
borders indicate that LL0 strongly screens (via $C_{01}$) the
inner LL1. In other words, most of the self-capacitance of LL1 is
taken up by capacitance to LL0. This screening of the inner LL is
ultimately what produces the chessboard pattern of the Kondo
conductance (see below). Furthermore, the length of the $N_1$
borders in Figs$.$ 2b,c alternates, as observed in the
experimental data (Fig$.$ 1b). The shorter $N_1$ borders appear as
only small discontinuities in the experimental Coulomb
oscillations. From Eq$.$ \ref{eq:double} it follows that this
alternating pattern results from the spin degeneracy of the FD
levels. Specifically, increasing $B$ and transferring a LL1
electron to LL0, at fixed $N$, costs an additional, LL0 level
spacing when $N_0$ changes from even to odd. The main feature of
the experimental data that is not present in Figs$.$ 2b,c is the
increased valley conductance due to the Kondo effect. We show
below, using our SDF analysis, that this increased valley
conductance appears in the hexagons where $N_0$ is odd.\\
\indent Our SDF calculations for realistic, 3D, lateral
semiconductor quantum dot structures have been thoroughly
described before \cite{Stopa96}. We compute self-consistent
eigenvalues $\varepsilon_{p \sigma}$, eigenfunctions $\psi_{p
\sigma}$, occupancies $n_{p \sigma}$ and tunneling coefficients
$\gamma_{p \sigma}$, with $p$ the orbital and $\sigma$ the spin
indices, as well as the total interacting energy $F$ of the
dot-gate-leads system, all as a function of $N$, $V_{gl}$ and $B$.
However, we cannot compute the coherent Kondo-assisted conductance
of the dot from the ground state properties provided by the SDF
calculation. Instead, we introduce an effective Kondo parameter,
which enables us to reproduce the conductance modulation
characteristic for the chessboard pattern.\\
\indent In Ref$.$ \cite{Keller01} it was qualitatively argued that
the alternating Kondo conductance with varying $B$ at fixed $N$
observed in their experiments in the strong edge state regime
resulted from Coulomb regulated redistribution of electrons, one
at a time, from LL1 to LL0. Since LL0 was assumed much more
strongly coupled to the leads, the Kondo effect was argued to
occur only when $N_0$ is odd. Here, the computed electronic
structure allows us to {\it quantitatively} exhibit this
depopulation process, to show how it varies in phase from one
Coulomb valley to the next and to show how the $B$-dependent
tunneling coefficients affect the structure of individual
chessboard fields. The model's assumption of spin-degenerate
states is accounted for at the end of the paper.\\
\indent In the standard derivation of the s-d model from the
Anderson model \cite{Hewson93}, a Schrieffer-Wolff transformation
is used to define an effective coupling between spin-degenerate
states of the dot (or impurity) via virtual electron exchange with
a neighboring Fermi sea. These co-tunneling processes then lead to
coherent hybridization of the degenerate states, which results in
a logarithmic increase of the conductance at low temperature, a
feature of the Kondo effect in quantum dots. In general, a dot can
contain many singly occupied electron states, for which spin-flip
can occur via co-tunneling contributing to the Kondo coupling. In
order to estimate the contribution of all singly-occupied states
to the Kondo effect, we define a parameter $K$, which we call
Kondo parameter, as the sum of all co-tunneling amplitudes that
leave the ground state of the system unchanged except for the flip
of a single spin
\begin{equation}
\begin{split}
K \equiv & \sum_{p, \sigma} n_{p, \sigma} (1- n_{p, \bar{\sigma}})
\sqrt{\gamma_{p, \sigma} \gamma_{p, \bar{\sigma}}} \\
& [ \frac{1}{F(N+1,V_{gl} B)-F(N,V_{gl},B)}+ \\
& \frac{1}{F(N-1,V_{gl},B)-F(N,V_{gl},B)}] \label{eq:K}
\end{split}
\end{equation}
where $\bar{\sigma}$ is the spin opposite to $\sigma$. We conclude
from the experimental data that the spin polarization of the leads
is negligible, in contrast to Ref$.$ \cite{Ciorga02}.\\
\indent A color scale plot of $K$ in the ($B$,$V_{gl}$) plane is
shown in Fig$.$ 3, for $N$= 32 to 39. Here, the $V_{gl}$
dependence has been approximated as follows. The two denominators
in Eq$.$ \ref{eq:K} are additions energies, which are, to a good
approximation, linear in $V_{gl}$ and vanish at the Coulomb
oscillations. Thus, $F(N+1,V_{gl},B)-F(N,V_{gl},B) \approx
(e^2/2C)+ e \alpha (V_{gl} - V_{gl}^{N,min}) + \varepsilon_{N+1}$,
where $(e^2/2C) \equiv (\partial^2 F/ \partial N^2)$ defines $C$,
and $\partial F (N,V_{gl}^{N,min},B)/ \partial N=0$ defines the
valley center gate voltage, $V_{gl}^{N,min}$, and where $\alpha =
C_{dot-gate}/C$ is the lever arm (a similar analysis holds for the
second energy denominator). By calculating the full electronic
structure only near the valley centers, we can determine the CI
parameters $V_{gl}^{N,min}$, $C$ and $\alpha$ and thereby show the
$V_{gl}$ dependence of $K$ due to the
 \begin{figure}[htbp]
   \begin{center}
     \centerline{\epsfig{file=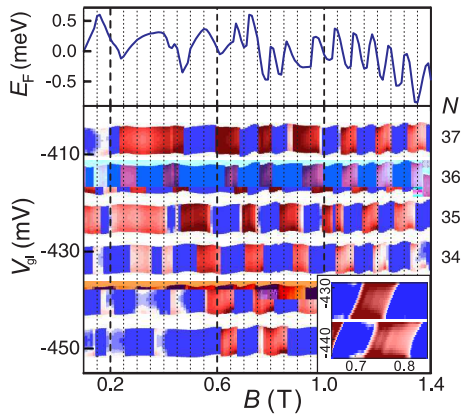, width=6cm, clip=true}}
     \caption{Color scale plot of Kondo parameter $K$ in the $B$,$V_{gl}$ plane.
     Red (blue) corresponds to large (small) $K$. Third LL filling begins below $ \sim 0.5 \, T$.
     $K$ is calculated in the Coulomb valley center at each $B$ (see text). Upper
     panel shows Fermi energy vs. $B$ for $N=37$ and fixed $V_{gl}$ = -407 mV.
     $E_F$ drops at each reconstruction (i$.$e$.$ at the end of each Kondo zone). Inset,
     $K$ fully calculated from Eq$.$ \ref{eq:K} on fine mesh for small $B,V_{gl}$ region (i$.$e$.$
     no approximation for the $V_{gl}$ dependence is used here).}
     \label{fig3}
   \end{center}
\vspace{-0.5cm}
 \end{figure}
\noindent addition energies, Fig$.$ 3. The full calculation of
Eq$.$ \ref{eq:K} on a mesh of $B,V_{gl}$ values, which is
numerically taxing, is shown for a small region in the inset to Fig$.$ 3.\\
\indent The alternating pattern of Kondo ``zones'' in successive Coulomb
valleys is evident. Even in this low magnetic field regime, the
coupling of the LL0 states to the leads typically exceeds that of
the LL1 states by two orders of magnitude. Therefore, although the
parameter $K$ is a sum of all possible co-tunneling amplitudes,
the amplitudes of the LL0 states dominate. Hence, even when $N$ is
odd, $K$ is negligible as long as $N_0$ is even.

Within a Kondo zone an abrupt increase of $K$ is followed by a
gradual decrease. This results from the contraction, with $B$, of
the half-filled orbit at the dot edge, and the resulting decrease
of its tunnel coefficient. When another electron depopulates from
LL1 to that orbit, that spin flip process is no longer available,
and $K$ collapses. Depopulation of LL1 coincides with a drop of
the dot's Fermi level \cite{EF} relative to the leads (Fig$.$ 3,
top panel). The denominators in Eq$.$ \ref{eq:K} are responsible
for the increase of $K$ away from the valley centers; a feature
which is clearly observed in the experiment Fig$.$ 1b. \\
\indent The approximation used in Fig$.$ 3 includes the $V_{gl}$
dependence of the charging energies, but not that of the $n_{p,
\sigma}$. The latter dependence, specifically that of the transfer
from LL1 to LL0, controls the slant of the chessboard fields. This
slant only emerges in the calculation of $K$ without
approximation, as shown for a small region of $V_{gl}$ and $B$
in the inset to Fig$.$ 3.\\
\indent Finally, we comment on our assumption of spin-degenerate
states. For $N$ odd, this only requires that the Zeeman splitting
($\sim$ 10 $\mu$eV) is negligible, which is the case in the $B$
range considered. However, for $N$ even, a singly filled orbital
in LL0 implies another one in LL1 and these electrons have, in
general, an exchange interaction. In Ref$.$ \cite{Tarucha00} it
was found that this exchange interaction diminishes rapidly with
$N$. For our dot, we explicitly determine the singlet-triplet
splitting for various (even) $N$ and $B$ by computing separate
ground states constrained to $S=0$ and $S=1$. We find that the
splitting is typically tens of $\mu$eV. The experimental signature
of a split ground state is a split Kondo resonance (in
source-drain voltage). Our analysis suggests that this splitting,
which has been observed \cite{Sasaki00,Fuhner02}, would be
characteristic of $N$ even in regimes where Zeeman energy is
small.\\
\indent We acknowledge financial support from the DARPA grant
number DAAD19-01-1-0659 of the QuIST program.

 \end{multicols}

\vspace{-0.5cm}
\begin{references}
\vspace{-1.5cm}
\bibitem{Kouwenhoven97} L.P. Kouwenhoven {\it et al.}, Electron transport in quantum dots,
in Mesoscopic Electron Transport, edited by L.L. Sohn {\it et
al.}, (Kluwer, Series E \textbf{345}, 1997), p.105-214.

\bibitem{Glazman88}  L.I. Glazman and M.E. Raikh, JETP Lett. \textbf{47}, 452 (1988).

\bibitem{Ng88}  T.K. Ng and P.A. Lee, Phys. Rev. Lett. \textbf{61}, 1768 (1988).

\bibitem{GG98}  D. Goldhaber-Gordon {\it et al.}, Nature \textbf{391}, 156 (1998).

\bibitem{Sara98}  S.M. Cronenwett {\it et al.}, Science \textbf{281}, 540 (1998).

\bibitem{Schmid98}  J. Schmid {\it et al.}, Physica B \textbf{256-258}, 182 (1998).

\bibitem{Maurer99}  S.M. Maurer {\it et al.}, Phys. Rev. Lett. \textbf{83}, 1403 (1999).

\bibitem{Wiel00} W.G. van der Wiel {\it et al.}, Science {\bf 289}, 2105
(2000).

\bibitem{Sasaki00}  S. Sasaki {\it et al.}, Nature \textbf{405}, 764
(2000).

\bibitem{Nygard00} J. Nyg\aa rd {\it et al.}, Nature \textbf{408}, 342 (2000).

\bibitem{Schmid00} J. Schmid {\it et al.}, Phys. Rev. Lett. \textbf{84}, 5824 (2000).

\bibitem{Keller01} M. Keller {\it et al.}, Phys. Rev. B \textbf{64}, 033302 (2001).

\bibitem{Wiel02} W.G. van der Wiel {\it et al.}, Phys. Rev. Lett. \textbf{88},
126803 (2002).

\bibitem{Sprinzak02} D. Sprinzak {\it et al.}, Phys. Rev. Lett. \textbf{88}, 176805 (2002).

\bibitem{Fuhner02} C. F\"{u}hner {\it et al.}, Phys. Rev. B \textbf{66}, 161305(R) (2002).

\bibitem{Keyser02} U.F. Keyser {\it et al.}, cond-mat/0206262 (2002).

\bibitem{Wiel03} W.G. van der Wiel {\it et al.}, Rev. Mod. Phys.
{\bf 75}(1), 1 (2003).

\bibitem{McEuen92} P.L. McEuen {\it et al.}, Phys. Rev. B \textbf{45}, 11419 (1992).

\bibitem{Kinaret93} J.M. Kinaret and N.S. Wingreen, Phys.
Rev. B \textbf{48}, 11113 (1993).

\bibitem{Vaart94} N.C. van der Vaart {\it et al.}, Phys.
Rev. Lett. \textbf{73}, 320 (1994).

\bibitem{FD}  V. Fock, Z. Phys. {\bf 47}, 446 (1928); C.G. Darwin, Proc.
Cambridge Philos. Soc. {\bf 27}, 86 (1930).

\bibitem{Stopa95} M. Stopa, Y. Aoyagi and T. Sugano, Phys. Rev. B {\bf 51}, 5494 (1995).

\bibitem{caps} A small upward adjustment to the capacitances,
by $< 50 \%$, is then made which appears to reproduce the
experimental data more faithfully. This might be attributed to
stronger lead-dot capacitance in the experimental structure than
that which is calculated.

\bibitem{Ciorga02} M. Ciorga {\it et al.}, Phys. Rev. Lett. \textbf{88}, 256804 (2002).

\bibitem{Stopa96} M. Stopa, Phys. Rev. B {\bf 54}, 13767 (1996);
M. Stopa, Semicond. Sci. Technol. {\bf 13}, A55 (1998); M. Stopa,
Physica E {\bf 10}, 103 (2001).

\bibitem{Hewson93} A.C. Hewson, {\it The Kondo Problem to Heavy
Fermions} (Cambridge, Cambridge, England, 1993).

\bibitem{Tarucha00} S. Tarucha {\it et al.}, Phys. Rev. Lett.
{\bf 84}, 2485 (2000).

\bibitem{EF} $E_F$ is defined by $N=\sum_i f[(\varepsilon_i -
E_F)/k_B T]$, with $\varepsilon_i$ Kohn-Sham levels, $N$ fixed and
$f$ the Fermi function.



\end{references}
\end{document}